\documentclass[preprint,12pt]{elsarticle}

\usepackage{amssymb}

\journal{Physica D: Nonlinear Phenomena}

\begin{document}

\begin{frontmatter}



\title{Canonical equations of Hamilton for the nonlinear Schr\"{o}dinger equation}


\author{Guo Liang}\author{Qi Guo\corref{cor1}}
\cortext[cor1]{Corresponding author.}
\ead{guoq@scnu.edu.cn}
\author{Zhanmei Ren}
\address{Laboratory of Nanophotonic Functional Materials and Devices, South
China Normal University, Guangzhou 510631}

\begin{abstract}
We define two different systems of mathematical physics: the second-order differential system (SODS) and the first-order differential system (FODS).
The Newton's second law of motion and the nonlinear Schr\"{o}dinger equation (NLSE) are the exemplary SODS and FODS, respectively.
We obtain a new kind of canonical equations of Hamilton (CEH), which are of some kind of symmetry in form and are formally different with the conventional CEH without symmetry [H. Goldstein, C. Poole,  J. Safko,
 Classical Mechanics, third ed., Addison-Wesley, 2001]. We also prove that the number of the CEHs is equal to the number of the generalized coordinates for the FODS, but twice the number of the generalized coordinates for the SODS. We show that the FODS can only be expressed by the new CEH,
but do not by the conventional CEH, while the SODS can be done by both the new and the conventional CEHs. As an example, we prove that the nonlinear Schr\"{o}dinger equation can be expressed with the new CEH in a consistent way.
\end{abstract}

\begin{keyword}
canonical equations of Hamilton \sep nonlinear Schr\"{o}dinger equation

\end{keyword}

\end{frontmatter}


\section{Introduction}  
The Hamiltonian viewpoint provides a framework for theoretical
extensions in many areas of
physics.
In classical mechanics it forms the basis for further developments,
such as Hamilton-Jacobi theory, perturbation approaches and chaos~\cite{Goldstein-book-05}. The canonical equations of Hamilton (CEH) in classical mechanics are expressed as~\cite{Goldstein-book-05}
\begin{eqnarray}
\dot{q_i}&=&\frac{\partial H}{\partial p_i},\quad (i=1,\cdots,n),\label{Hamilton formulations 1 for discrete system}\\
 -\dot{p_i}&=&\frac{\partial H}{\partial q_i},\quad (i=1,\cdots,n),\label{Hamilton formulations 2 for discrete system}
\end{eqnarray}
where $q_i$ and $p_i$ are said to be the generalized coordinate and the generalized momentum, $\dot{q}_i=d q_i/dt$ is the generalized velocity, $\dot{p}_i=d p_i/dt$, and $H$ is the Hamiltonian of the system. The generalized momentum $p_i$ is defined as
$p_i=\frac{\partial L}{\partial\dot{q}_i}$, where $L$ is the Lagrangian of the system. The Hamiltonian $H$ is obtained by the Legendre transformation
$H=\sum_{i=1}^n\dot{q}_ip_i-L$.
 A set of $(q_i,p_i)$ forms a $2n$-dimensional phase space.

The CEH (\ref{Hamilton formulations 1 for discrete system}) and (\ref{Hamilton formulations 2 for discrete system}) are not only valid for the discrete system, but also can be extended to the continuous system as~\cite{Goldstein-book-05}
\begin{eqnarray}
\dot{q}_s&=&\frac{\delta h}{\delta p_s},\quad (s=1,\cdots,N),\label{derivative of q for continuous system}\\
-\dot{p}_s&=&\frac{\delta h}{\delta q_s},\quad (s=1,\cdots,N),\label{derivative of
pi for continuous system}
\end{eqnarray}
where the subscript $s$, which all denotes $1,\cdots,N$  throughout the paper, represents the components of the quantity of the continuous system~\cite{Goldstein-book-05}, $\frac{\delta h}{\delta q_s}=\frac{\partial h}{\partial q_s}-\frac{\partial}{\partial x}\frac{\partial h}{\partial q_{s,x}}$ and $\frac{\delta h}{\delta p_s}=\frac{\partial h}{\partial p_s}-\frac{\partial}{\partial x}\frac{\partial h}{\partial p_{s,x}}$ denote the functional derivatives of $h$ with respect to $q_s$ and $p_s$ with $q_{s,x}=\frac{\partial q_s}{\partial x}$ and $p_{s,x}=\frac{\partial p_s}{\partial x}$, $q_s$ and $p_s$ are the generalized coordinate and the generalized momentum, respectively, and $h$ is the Hamiltonian density of the continuous system. Like the discrete system, the generalized momentum $p_s$ for the continuous system is defined as
 \begin{equation}\label{conjugate momenta}
 p_s=\frac{\partial l}{\partial\dot{q}_s},
 \end{equation}
 and the Hamiltonian density $h$ for the continuous system is obtained by the Legendre transformation as
 \begin{equation}\label{legendre transform}
 h=\sum_{s=1}^N\dot{q}_sp_s-l,
 \end{equation}
where $l$ is the Lagrangian density.
But it is significantly different for the continuous system that $q_s$ and $p_s$ are now not only functions of time $t$, but also the spatial coordinate $x$, where the spatial coordinate $x$
is not the generalized coordinate, but only serves as the continuous index replacing the discrete $i$
in Eqs.(\ref{Hamilton formulations 1 for discrete system}) and (\ref{Hamilton formulations 2 for discrete system}). To distinguish $t$ from the coordinate $x$, we refer to time $t$
as the evolution coordinate. $p_s$ and $q_s$ define the infinite-dimensional phase space.
$h$ is a function of $q_s,p_s$ and $q_{s,x}$ but not $p_{s,x}$~\cite{Goldstein-book-05}, so $\frac{\delta h}{\delta p_s}=\frac{\partial h}{\partial p_s}$, then Eq.(\ref{derivative of q for continuous system}) can be also expressed as
\begin{equation}\label{another form}
\dot{q}_s=\frac{\partial h}{\partial p_s}.
\end{equation}

``The advantages of the Hamiltonian formulation lie not
in its use as a calculational tool," as has been pointed out in Ref.\cite{Goldstein-book-05}, ``but rather in the deeper insight it affords into the formal structure of mechanics."
We have a greater freedom to select the physical quantities to designate as ``coordinates'' and ``momenta'', which have equal status as independent variables.
  As a result we can present the physical content of mechanics in a newer and more abstract way. The more abstract formulation is primarily of interest to us today because of its essential role in constructing the framework of modern physics, such as statistical mechanics and quantum theory.

  Up to now, to our knowledge, the CEH appearing in all the literatures are of the form (\ref{derivative of q for continuous system})
  and (\ref{derivative of pi for continuous system}), which are established based on the the second-order differential system (SODS)
  (the definition of the SODS will be given in the following). Besides the SODS, there are a number of the first-order differential
  systems (FODS) (the definition of the FODS will also be given in the following) to model physical phenomena. For example, the nonlinear Schr\"{o}dinger equation (NLSE) is a universal FODS.
  A question is raised in the nature of things: what is the form of the CEH valid for the FODS? Are the conventional CEH, Eqs. (\ref{derivative of q for continuous system}) and (\ref{derivative of pi for continuous system}), still valid for the FODS? In Ref.\cite{Hasegawa-book-95}, the CEH
for the NLSE were considered to be the same as Eqs.(\ref{derivative of q for continuous system}) and (\ref{derivative of pi for continuous system}). But, as will be shown in the paper, the conventional CEH, Eqs. (\ref{derivative of q for continuous system}) and (\ref{derivative of pi for continuous system}), are not valid for the NLSE, from which it is impossible to obtain the NLSE or its complex-conjugate equation. Attempt was made to deal with the difficulty in Ref.\cite{Dauxois-book-06},
but the CEH for the NLSE they obtained were inconsistent, as will be shown later. In this paper, we obtain a new kind of CEH valid
for the FODS, with which the NLSE can be expressed in a consistent way. We prove that the new CEH
 and the conventional CEH 
 are equivalent for the description of the SODS. But the conventional CEH
 can not express the FODS.


The paper is organized as follows. We define the SODS and the FODS, and obtain the new CEH with symmetry valid for the FODS in Section \ref{CEH for first order}. The application of the new CEH obtained in Section \ref{CEH for first order} to the NLSE is discussed in Section \ref{CEH for NLSE}. The NLSE can be expressed with the new CEH in a consistent way, but can not be expressed with the conventional CEH. The further discussion about the new CEH with symmetry is in Section \ref{Relation}. We find that the new CEH with symmetry obtained in Section \ref{CEH for first order} can also be used to express the SODS. We also prove that the number of the CEHs is equal to the number of the generalized coordinates for the FODS, but twice the number of the generalized coordinates for the SODS. Section \ref{conclusion} gives a summary.

\section{First-order differential system and its canonical equations of Hamilton}\label{CEH for first order}

The Newton's second law of motion in classical mechanics, based on which the Hamiltonian formulation is established, is the second-order differential equation about the evolution coordinate (here the evolution coordinate is time). In this paper we define the system described by the second-order partial differential equation about the evolution coordinate as the second-order differential system (SODS). Similarly, the first-order differential system (FODS) is the system described by the first-order partial differential equation about the evolution coordinate. The Lagrangian density of the SODS of the continuous systems is expressed in general as ~\cite{Goldstein-book-05}
\begin{equation}\label{Lagrangian density of 2nd system}
l=\sum_{s=1}^N\sum_{k=1}^NA_{sk}\dot{q}_s\dot{q}_k+\sum_{s=1}^NB_s\dot{q}_s+C,
\end{equation}
where $A_{sk},B_{s},C$ depend on not only $q_s$ but also $q_{s,x}$ in general.
The generalized momentum can be obtained by the definition (\ref{conjugate momenta}) as
\begin{equation}
p_s=\sum_{k=1}^N(A_{sk}+A_{ks})\dot{q}_k+B_s\label{conjugate momenta 2},
\end{equation}
which is a function of $q_s$, $\dot{q}_s$ and $q_{s,x}$.
The number of Eqs. (\ref{conjugate momenta 2}) is $N$, and there are $4N$ variables, which are $q_s$, $\dot{q}_s$, $p_s$ and $q_{s,x}$. So the degree of freedom of  Eqs. (\ref{conjugate momenta 2}) is $3N$. Then $q_s,p_s$ and $q_{s,x}$ are taken as independent variables. The generalized velocities $\dot{q}_s$ can be expressed with these independent variables.

Besides the SODSs, there are a number of the FODSs to model physical phenomena, for example, the nonlinear Schr\"{o}dinger equation (NLSE)
\begin{equation}\label{NNLSE}
i\frac{\partial \varphi}{\partial
t}+\frac{1}{2}\frac{\partial^2\varphi}{\partial x^2}+|\varphi|^2\varphi=0,
\end{equation}
which is
a universal
model that can be
applied to hydrodynamics~\cite{Nore-PhysicaD-93}, nonlinear
optics~\cite{Haus-book,Hasegawa-book-95,Agrawal-book}, nonlinear
acoustics~\cite{Bisyarin-AIPConf.Proc.-08}, Bose-Einstein
condensates~\cite{seaman-pra-05}, and can also describe waves on deep water~\cite{Zakharov-jamtp-1968} and Langmuir waves in plasma~\cite{Akhmediev-book-2005}. In nonlinear optics~\cite{Hasegawa-book-95,Haus-book,Agrawal-book}, the NLSE (\ref{NNLSE}) governs the propagation of the slowly-varying light-envelope,
where the evolution coordinate $t$ is the propagation
direction coordinate, the light-envelope $\varphi$ is a cw paraxial beam
in a planar waveguide~\cite{Haus-book} or a narrow spectral-width pulse
in optical fibers~\cite{Hasegawa-book-95,Agrawal-book}, and $x$ is a transverse space coordinate for the beam and a frame moving at the group velocity (the so-called retarded frame)
for the pulse, respectively.

For the FODS,
the Lagrangian density must be the linear function of the generalized velocities $\dot{q}_s$.
 If the Lagrangian density is a quadratic function of the generalized velocities like Eq.(\ref{Lagrangian density of 2nd system}), the
equation of motion, i.e., the Euler-Lagrange equation
 \begin{equation}\label{Euler-Lagrange equation}
\frac{\partial}{\partial t}\frac{\partial l}{\partial\dot{q}_s}-\frac{\delta l}{\delta q_s}=0
\end{equation}
will be the second-order partial differential equation about the evolution coordinate $t$, which is
 in contradiction with the definition of the FODS. Therefore, the Lagrangian density of the FODS can only be expressed as
\begin{equation}\label{Lagrangian density for the fist-order system}
l=\sum_{s=1}^NR_s(q_s)\dot{q}_s+Q(q_s,q_{s,x}).
\end{equation}
Besides, $R_s$ in Eq.(\ref{Lagrangian density for the fist-order system}) is not the function of a set of $q_{s,x}$. If $R_s$ is also the function of a set of $q_{s,x}$, there will be such terms as $q_{s,x}\dot{q}_s$ appearing in Eq.(\ref{Lagrangian density for the fist-order system}). Substitution of Eq.(\ref{Lagrangian density for the fist-order system}) into Eq.(\ref{Euler-Lagrange equation}) leads to the appearance of the mixed partial derivative terms $\frac{\partial^2 q_s}{\partial x\partial t}$. Via the coordinate rotation transform, terms $\frac{\partial^2 q_s}{\partial t^2}$ and $\frac{\partial^2 q_s}{\partial x^2}$ will appear instead. Therefore,
the Euler-Lagrange equation (\ref{Euler-Lagrange equation}) expressed with the canonical form of the second-order partial differential equation~\cite{Arfken-book-2005} is the second-order partial differential equation about the evolution coordinate $t$. According to our definition,
 the system
 is the SODS. Consequently, the generalized momentum $p_s$, which is obtained by the definition (\ref{conjugate momenta}) as
\begin{equation}\label{equations of p q}
p_s=R_s(q_s),
\end{equation}
is only a function of $q_s$. This is of significant difference from the case of the SODS, where the generalized momentum $p_s$ is the function of not only $q_s$, but also $\dot{q}_s$ and $q_{s,x}$, as shown in Eq.(\ref{conjugate momenta 2}). There are $2N$ variables, $q_s$ and $p_s$, in Eqs. (\ref{equations of p q}). The number of Eqs. (\ref{equations of p q}) is $N$, which also means there exist $N$ constraints between $q_s$ and $p_s$. So the degree of freedom of the system given by Eqs. (\ref{equations of p q}) is $N$. Without loss of generality, we take $q_1,\cdots,q_\nu$ and $p_1,\cdots,p_\mu$ as the independent variables, where $\nu+\mu=N$. The remaining generalized coordinates and generalized momenta can be expressed with these independent variables as
$
q_\alpha=f_\alpha(q_1,\cdots,q_\nu,p_1,\cdots,p_\mu)(\alpha=\nu+1,\cdots,N),$ and $
p_\beta=g_\beta(q_1,\cdots,q_\nu,p_1,\cdots,p_\mu) (\beta=\mu+1,\cdots,N).
$

We now derive the CEH for the FODS.
 The total differential of the Hamiltonian density $h$ can be obtained by
using Eq.(\ref{legendre transform})
{\setlength\arraycolsep{0pt}
\begin{equation}
d h=\sum_{s=1}^Np_sd \dot{q}_s+\sum_{\eta=1}^{\mu}\dot{q}_\eta d p_\eta+\sum_{\beta=\mu+1}^{N}\dot{q}_\beta\left(\sum_{\lambda=1}^{\nu}\frac{\partial g_\beta}{\partial q_\lambda}d q_\lambda+\sum_{\eta=1}^{\mu}\frac{\partial g_\beta}{\partial p_\eta}d p_\eta\right)
-d l,\label{eq1}
\end{equation}}
while the total differential of the Lagrangian density $l(q_s,\dot{q}_s,q_{s,x})$ with respect to its arguments is
{\setlength\arraycolsep{0pt}
\begin{equation}
d l=\sum_{\lambda=1}^{\nu}\frac{\partial l}{\partial q_\lambda}d q_\lambda+\sum_{\alpha=\nu+1}^{N}\frac{\partial l}{\partial q_\alpha}\left(\sum_{\lambda=1}^{\nu}\frac{\partial f_\alpha}{\partial q_\lambda}d q_\lambda+\sum_{\eta=1}^{\mu}\frac{\partial f_\alpha}{\partial p_\eta}d p_\eta\right)+\sum_{s=1}^N\frac{\partial l}{\partial\dot{q}_s}d \dot{q}_s+\sum_{s=1}^N\frac{\partial l}{\partial q_{s,x}}d q_{s,x}.\label{eq2}
\end{equation}}
 Substitution of Eq.(\ref{eq2}) into Eq.(\ref{eq1}) yields
 {\setlength\arraycolsep{0pt}
\begin{eqnarray}
d h&=&\sum_{\lambda=1}^{\nu}\left(\sum_{\beta=\mu+1}^{N}\dot{q}_\beta\frac{\partial g_\beta}{\partial q_\lambda}-\sum_{\alpha=\nu+1}^{N}\frac{\partial l}{\partial q_\alpha}\frac{\partial f_\alpha}{\partial q_\lambda}-\frac{\partial l}{\partial q_\lambda}\right)d q_\lambda\nonumber\\
&&+\sum_{\eta=1}^{\mu}\left(\dot{q}_\eta+\sum_{\beta=\mu+1}^{N}\dot{q}_\beta\frac{\partial g_\beta}{\partial p_\eta}-\sum_{\alpha=\nu+1}^{N}\frac{\partial l}{\partial q_\alpha}\frac{\partial f_\alpha}{\partial p_\eta}\right)d p_\eta\nonumber\\
&&-\sum_{s=1}^N\frac{\partial l}{\partial q_{s,x}}d q_{s,x}.\label{differential equations of H1}
\end{eqnarray}}
Since the total differential of $h(q_1,\cdots,q_\nu,p_1,\cdots,p_\mu,q_{s,x})$ with respect to its arguments can be written as
$
d h=\sum_{\lambda=1}^{\nu}\frac{\partial h}{\partial q_\lambda}d q_\lambda+\sum_{\eta=1}^{\mu}\frac{\partial h}{\partial p_\eta}d p_\eta+\sum_{s=1}^N\frac{\partial h}{\partial q_{s,x}}d q_{s,x},
$
by comparing this equation with Eq.(\ref{differential equations of H1}), we obtain $2N$ equations
{\setlength\arraycolsep{0pt}
\begin{eqnarray}
\frac{\partial h}{\partial q_\lambda}&=&\sum_{\beta=\mu+1}^{N}\dot{q}_\beta\frac{\partial g_\beta}{\partial q_\lambda}-\sum_{\alpha=\nu+1}^{N}\frac{\partial l}{\partial q_\alpha}\frac{\partial f_\alpha}{\partial q_\lambda}-\frac{\partial l}{\partial q_\lambda},\label{h q}\\
\frac{\partial h}{\partial p_\eta}&=&\dot{q}_\eta+\sum_{\beta=\mu+1}^{N}\dot{q}_\beta\frac{\partial g_\beta}{\partial p_\eta}-\sum_{\alpha=\nu+1}^{N}\frac{\partial l}{\partial q_\alpha}\frac{\partial f_\alpha}{\partial p_\eta},\label{h p}\\
\frac{\partial h}{\partial q_{s,x}}&=&-\frac{\partial l}{\partial q_{s,x}},\label{h qx}
\end{eqnarray}}
where $\lambda=1,\cdots,\nu,\eta=1,\cdots,\mu,s=1,\cdots,N$. From the Euler-Lagrange equations (\ref{Euler-Lagrange equation}),
we obtain
\begin{equation}\label{Euler-Lagrange equation1}
\frac{\partial l}{\partial q_s}=\frac{\partial}{\partial t}\frac{\partial l}{\partial\dot{q}_s}+\frac{\partial}{\partial x}\frac{\partial l}{\partial q_{s,x}}.
\end{equation}
Substituting Eq.(\ref{Euler-Lagrange equation1}) into Eqs.(\ref{h q}) and (\ref{h p}), we have
{\setlength\arraycolsep{0pt}
\begin{eqnarray}
\frac{\partial h}{\partial q_\lambda}&=&-\dot{p}_\lambda+\sum_{\beta=\mu+1}^{N}\dot{q}_\beta\frac{\partial g_\beta}{\partial q_\lambda}-\sum_{\alpha=\nu+1}^{N}\dot{p}_\alpha\frac{\partial f_\alpha}{\partial q_\lambda}-\sum_{\alpha=\nu+1}^{N}\frac{\partial}{\partial x}\frac{\partial l}{\partial q_{\alpha,x}}\frac{\partial f_\alpha}{\partial q_\lambda}-\frac{\partial}{\partial x}\frac{\partial l}{\partial q_{\lambda,x}},\label{h q-2}\\
\frac{\partial h}{\partial p_\eta}&=&\dot{q}_\eta+\sum_{\beta=\mu+1}^{N}\dot{q}_\beta\frac{\partial g_\beta}{\partial p_\eta}-\sum_{\alpha=\nu+1}^{N}\dot{p}_\alpha\frac{\partial f_\alpha}{\partial p_\eta}-\sum_{\alpha=\nu+1}^{N}\frac{\partial}{\partial x}\frac{\partial l}{\partial q_{\alpha,x}}\frac{\partial f_\alpha}{\partial p_\eta}.\label{h p-2}
\end{eqnarray}}
Then substituting Eq.(\ref{h qx}) into Eqs.(\ref{h q-2}) and (\ref{h p-2}), we obtain $N$ CEHs for the FODS
{\setlength\arraycolsep{0pt}
\begin{eqnarray}
\frac{\delta h}{\delta q_\lambda}&=&-\dot{p}_\lambda+\sum_{\beta=\mu+1}^{N}\dot{q}_\beta\frac{\partial g_\beta}{\partial q_\lambda}-\sum_{\alpha=\nu+1}^{N}\dot{p}_\alpha\frac{\partial f_\alpha}{\partial q_\lambda}+\sum_{\alpha=\nu+1}^{N}\frac{\partial}{\partial x}\frac{\partial h}{\partial q_{\alpha,x}}\frac{\partial f_\alpha}{\partial q_\lambda},\label{h q-last}\\
\frac{\delta h}{\delta p_\eta}&=&\dot{q}_\eta+\sum_{\beta=\mu+1}^{N}\dot{q}_\beta\frac{\partial g_\beta}{\partial p_\eta}-\sum_{\alpha=\nu+1}^{N}\dot{p}_\alpha\frac{\partial f_\alpha}{\partial p_\eta}+\sum_{\alpha=\nu+1}^{N}\frac{\partial}{\partial x}\frac{\partial h}{\partial q_{\alpha,x}}\frac{\partial f_\alpha}{\partial p_\eta}.\label{h p-last}
\end{eqnarray}}
To obtain Eq.(\ref{h p-last}), we have used $\frac{\delta h}{\delta p_\eta}=\frac{\partial h}{\partial p_\eta}$, because $h$ is not a function of $p_{\eta,x}$.
The CEH, Eqs. (\ref{h q-last}) and (\ref{h p-last}), can be expressed in a symmetric form as
{\setlength\arraycolsep{0pt}
\begin{eqnarray}
\frac{\delta h}{\delta q_\lambda}&=&\sum_{s=1}^N\left(\dot{q}_s\frac{\partial p_s}{\partial q_\lambda}-\dot{p}_s\frac{\partial q_s}{\partial q_\lambda}\right)+\sum_{\alpha=\nu+1}^{N}\frac{\partial}{\partial x}\frac{\partial h}{\partial q_{\alpha,x}}\frac{\partial f_\alpha}{\partial q_\lambda},\label{canonical equations 1 for 1st order }\\
\frac{\delta h}{\delta p_\eta}&=&\sum_{s=1}^N\left(\dot{q}_s\frac{\partial p_s}{\partial p_\eta}-\dot{p}_s\frac{\partial q_s}{\partial p_\eta}\right)+\sum_{\alpha=\nu+1}^{N}\frac{\partial}{\partial x}\frac{\partial h}{\partial q_{\alpha,x}}\frac{\partial f_\alpha}{\partial p_\eta}\label{canonical equations 2 for 1st order }
\end{eqnarray}
}($\lambda=1,\cdots,\nu$, $\eta=1,\cdots,\mu$, and $\nu+\mu=N$),
because $\dot{p}_\lambda=\sum_{\lambda'=1}^\nu\dot{p}_{\lambda'}\frac{\partial q_{\lambda'}}{\partial q_\lambda},  \dot{q}_\eta=\sum_{\eta'=1}^\mu\dot{q}_{\eta'}\frac{\partial p_{\eta'}}{\partial p_\eta}$, $\sum_{\eta=1}^{\mu}\dot{q}_\eta\frac{\partial p_\eta}{\partial q_\lambda}=0,$ and $\sum_{\lambda=1}^{\nu}\dot{p}_\lambda\frac{\partial q_\lambda}{\partial p_\eta}=0$.
The CEH, Eqs. (\ref{canonical equations 1 for 1st order }) and (\ref{canonical equations 2 for 1st order }), can be easily extended to the discrete system, which can be expressed as
{\setlength\arraycolsep{0pt}
\begin{eqnarray}
\frac{\partial H}{\partial q_\lambda}&=&\sum_{s=1}^N\left(\dot{q}_s\frac{\partial p_s}{\partial q_\lambda}-\dot{p}_s\frac{\partial q_s}{\partial q_\lambda}\right),\label{canonical equations 1 for 1st order for discrete system }\\
\frac{\partial H}{\partial p_\eta}&=&\sum_{s=1}^N\left(\dot{q}_s\frac{\partial p_s}{\partial p_\eta}-\dot{p}_s\frac{\partial q_s}{\partial p_\eta}\right),\label{canonical equations 2 for 1st order for discrete system}
\end{eqnarray}
}
where $\lambda=1,\cdots,\nu$, $\eta=1,\cdots,\mu$, and $\nu+\mu=N$.

\section{Application of the new canonical equations of Hamilton with symmetry to the nonlinear Schr\"{o}dinger equation}\label{CEH for NLSE}
In this part, we will discuss the application of the new CEH with symmetry, Eqs. (\ref{canonical equations 1 for 1st order }) and (\ref{canonical equations 2 for 1st order }), to the NLSE. It is known that the Lagrangian density for the NLSE can be
expressed as~\cite{Anderson-pra-83}
$
l=-\frac{i}{2}(\varphi^*\frac{\partial\varphi}{\partial
t}-\varphi\frac{\partial \varphi^*}{\partial
t})+\frac{1}{2}|\frac{\partial\varphi}{\partial x}|^2-\frac{1}{2}|\varphi|^4
$. The NLSE is complex, and therefore it is an equation with two real functions, the real part of $\varphi$ and its imaginary part. It is convenient to consider
instead the fields $\varphi$ and $\varphi^*$ which are treated as independent from each other. Therefore, $N$  (the components of the quantity) for the NLSE equals two,
i.e., there are two generalized coordinates, $q_1=\varphi^*$ and $q_2=\varphi$, and two generalized momenta can be obtained by using the definition (\ref{conjugate momenta}) as
\begin{equation}\label{generalized momenta for NLSE}
p_1=\frac{i}{2}\varphi,
p_2=-\frac{i}{2}\varphi^*.
\end{equation}
The Hamiltonian density for the NLSE can be obtained by use of
Eq.(\ref{legendre transform}) as \cite{Hasegawa-book-95}
\begin{equation}\label{Hamiltonian density for NLSE}
h=-\frac{1}{2}\left|\frac{\partial\varphi}{\partial x}\right|^2+
\frac{1}{2}|\varphi|^4.
\end{equation}
If the generalized coordinate $q_1$ and the generalized momentum $p_1$ are taken as the independent variables, the remaining generalized coordinate $q_2$ and the remaining generalized momentum $p_2$ can be expressed via the relations (\ref{generalized momenta for NLSE}) as $q_2=-2ip_1$ and $p_2=-\frac{i}{2}q_1$, respectively. We should also note that the Hamiltonian density $h$ is also the function of $q_{s,x}$, which are independent from $q_s$ and $p_s$. Then for the NLSE, the Hamiltonian density (\ref{Hamiltonian density for NLSE}) should be expressed with the independent variables $q_1,p_1,q_{1,x}$ and $q_{2,x}$ as
\begin{equation}\label{Hamiltonian density for NLSE with independent variables}
h=-\frac{1}{2}q_{1,x}q_{2,x}-2q_1^2p_1^2.
\end{equation}
We should note that $\nu=\mu=1$ and $N=2$ for the NLSE, which means that the equations (\ref{canonical equations 1 for 1st order }) have only one equation, so do Eqs.(\ref{canonical equations 2 for 1st order }). Therefore, the CEH (\ref{canonical equations 1 for 1st order }) and (\ref{canonical equations 2 for 1st order }) will yield two equations for the NLSE.
The left side of Eq.(\ref{canonical equations 1 for 1st order })
is obtained as
\begin{equation}
\frac{\delta h}{\delta q_1}=\frac{1}{2}\frac{\partial^2 \varphi}{\partial x^2}+|\varphi|^2\varphi,
\end{equation} and
its right side is
\begin{equation}
-\dot{p}_1+\dot{q}_2\frac{\partial p_2}{\partial q_1}=-i\dot{\varphi}.
\end{equation}
Then the NLSE (\ref{NNLSE}) can be obtained.
Using the other CEH, the left side of Eq.(\ref{canonical equations 2 for 1st order }) is obtained as
\begin{equation}\label{derivative of h to varphi}
\frac{\delta h}{\delta p_2}=-4q_1^2p_1=-2i|\varphi|^2\varphi^*,
\end{equation}
and its right side is
\begin{equation}
\dot{q}_1-\dot{p}_2\frac{\partial q_2}{\partial p_1}+\frac{\partial}{\partial x}\frac{\partial h}{\partial q_{2,x}}\frac{\partial q_2}{\partial p_1}=2\dot{\varphi}^*+i\frac{\partial^2\varphi^*}{\partial x^2}.
\end{equation}
Then the complex conjugate of the NLSE is also obtained. Therefore, the CEHs (\ref{canonical equations 1 for 1st order }) and (\ref{canonical equations 2 for 1st order }) are consistent in the sense that the NLSE can be expressed with one of the two CEHs,
and its complex conjugate can be expressed with the other.

In Ref.\cite{Hasegawa-book-95}, the CEH for the NLSE were considered to be the same as those for the SODS, that is, Eqs. (\ref{derivative of q for continuous system})
and (\ref{derivative of pi for continuous system}).
Then, according to the definition~(\ref{conjugate momenta}), $p_\varphi=\partial l/\partial\dot{\varphi}$,
$p_\varphi$ must be $-i/2\varphi^*$ but not $-i\varphi^*$.
It was artificially doubled in Ref.\cite{Hasegawa-book-95} so that $p_\varphi=-i\varphi^*$
in Eq.[5.1.29] (to avoid confusion, we replace the parentheses by the brackets to represent the formulas in the references) to make the NLSE
derived from the CEH (\ref{derivative of pi for continuous system}).
In fact, substitution of the Hamiltonian density (\ref{Hamiltonian density for NLSE}) into Eq.(\ref{derivative of pi for continuous system}) only yields $
\frac{i}{2}\frac{\partial \varphi}{\partial
t}+\frac{1}{2}\nabla_\bot^2\varphi+|\varphi|^2\varphi=0,
$
which in fact does not be the NLSE (\ref{NNLSE}). In Ref.\cite{Dauxois-book-06}, the CEH obtained by the authors are Eqs.[3.87] and [3.81], the latter is the same as Eq.(\ref{another form}).
Although the NLSE can be derived from Eq.[3.87], its complex conjugation could not be obtained from the other, Eq.~[3.81]. We now show this claim.
 Substituting the Hamiltonian density (\ref{Hamiltonian density for NLSE}) into Eq.(\ref{another form}), the left side of it is obtained as $\frac{\partial \varphi^*}{\partial t}$, and the right side is $\frac{\partial h}{\partial p_{\varphi^*}}=\frac{\partial h}{\partial \varphi}\frac{\partial \varphi}{\partial p_{\varphi^*}}=-2i|\varphi|^2\varphi^*$, where $p_{\varphi^*}=\frac{\partial l}{\partial\dot{\varphi}^*}$. Then the equation
$
-\frac{i}{2}\frac{\partial \varphi^*}{\partial
t}+\left|\varphi\right|^2\varphi^*=0
$
can be obtained,
which is absolutely not the complex conjugate of the NLSE. Therefore, the CEH for the NLSE obtained in Ref.\cite{Dauxois-book-06} are inconsistent.

\section{Some further discussions about the new canonical equations of Hamilton with symmetry}\label{Relation}
Now, we know that the CEH valid for the FODS are Eqs.~(\ref{canonical equations 1 for 1st order }) and (\ref{canonical equations 2 for 1st order }), and the conventional CEH valid for the SODS are Eqs.~(\ref{derivative of q for continuous system})
and (\ref{derivative of pi for continuous system}). We will prove here that the new CEH with symmetry, Eqs. (\ref{canonical equations 1 for 1st order }) and (\ref{canonical equations 2 for 1st order }), can also be used to express the SODS.
If all the $N$ generalized coordinates $q_s$ and $N$ generalized momenta $p_s$ in Eqs.(\ref{canonical equations 1 for 1st order })
and (\ref{canonical equations 2 for 1st order }) are independent, we obtain $\frac{\partial g_\beta}{\partial q_\lambda}=
\frac{\partial f_\alpha}{\partial q_\lambda}=\frac{\partial g_\beta}{\partial p_\eta}=
\frac{\partial f_\alpha}{\partial p_\eta}=0$ ($\alpha=\nu+1,\cdots,N,\beta=\mu+1,\cdots,N$), then the CEH, Eqs. (\ref{canonical equations 1 for 1st order })
and (\ref{canonical equations 2 for 1st order }), can be reduced to Eqs.(\ref{derivative of q for continuous system})
and (\ref{derivative of pi for continuous system}).
 In fact, this is just the case of the SODS, where all the generalized coordinates and the generalized momenta are independent. Consequently, the new CEH with symmetry obtained in the paper can express not only the FODS but also the SODS. In the other word, the new CEH, Eqs. (\ref{canonical equations 1 for 1st order })
and (\ref{canonical equations 2 for 1st order }), and the conventional CEH, Eqs.(\ref{derivative of q for continuous system})
and (\ref{derivative of pi for continuous system}), are equivalent for the description of the SODS, but the former are with some kind of symmetry in form and the latter are lack of such symmetry. The conventional CEH, Eqs.(\ref{derivative of q for continuous system})
and (\ref{derivative of pi for continuous system}), can only expresses the SODS.

In addition, the number of Eqs. (\ref{canonical equations 1 for 1st order })
and (\ref{canonical equations 2 for 1st order }) is $N$ ($N$ is the number of the generalized coordinates) in the case for the FODS, and is half of that of Eqs.(\ref{derivative of q for continuous system}) and (\ref{derivative of pi for continuous system}).
This can be explained in the following way. For the SODS, the Euler-Lagrange equations (\ref{Euler-Lagrange equation})
 are the second-order partial differential equations about the evolution coordinate, where the Lagrangian density is expressed as Eq.(\ref{Lagrangian density of 2nd system}). The number of Eqs.(\ref{Euler-Lagrange equation}) is $N$. It is well known that one second-order differential equation can be reduced to two first-order differential equations~\cite{M.Lea-book-2004}. Then $2N$ CEHs, which are the first-order partial differential equations about the evolution coordinate, can be obtained from the $N$ Euler-Lagrange equations. But it is significantly different for the FODS that the Euler-Lagrange equations (\ref{Euler-Lagrange equation}) are the first-order partial differential equations about the evolution coordinate, so only $N$ CEHs are obtained from Eqs.(\ref{Euler-Lagrange equation}).

\section{Conclusion}\label{conclusion}
We obtain a new kind of CEH, which are of some kind
of symmetry in form. The new CEH with symmetry can express both the FODS and the SODS, while the conventional CEH can only express the SODS but impossibly express the FODS. The number of the CEH for the first-order differential
system is $N$ rather than
$2N$ like the case for the second-order differential
system, where $N$ is the number of the generalized coordinates. The NLSE can be expressed with the new CEH in a consistent way, but can not be expressed with
the conventional CEH. The CEH for the NLSE are two equations, which are consistent in the sense that the NLSE can be expressed with one of the CEH and its complex conjugate can be expressed with the other. The Hamiltonian formulation for the continuous system can also be extended to the discrete system.

\section*{Acknowledgment}

This research was supported by the National Natural Science
 Foundation of China (Grant Nos. 11274125, 11074080), and the Natural Science Foundation of Guangdong Province of China (Grant No. S2012010009178).





\bibliographystyle{model1a-num-names}
\bibliography{<your-bib-database>}



\end{document}